\documentclass[aps,prl,superscriptaddress,showpacs,nofootinbib,floatfix,twocolumn]{revtex4}

\usepackage{doi,graphicx,amssymb,amsmath,dcolumn,gensymb}

\newcommand{\apjs}{\textrm{ApJS}}

\newcommand{\aap}{\textrm{A\&A}}

\newcolumntype{d}[1]{D{.}{.}{#1} }
\usepackage[percent]{overpic}

\begin{document}

\title{Nuclear pasta and supernova neutrinos at late times}

\author{C. J. Horowitz}\email{horowit@indiana.edu}
\author{D. K. Berry}
\author{M. E. Caplan}
\affiliation{Center for Exploration of Energy and Matter and Department of Physics, Indiana University, Bloomington, IN 47405, USA}
\author{T. Fischer} 
\affiliation{Institute for Theoretical Physics, University of Wroclaw, plac Maksa Borna 9, 50-204 Wroclaw, Poland}
\author{Zidu Lin}
\affiliation{Center for Exploration of Energy and Matter and Department of Physics, Indiana University, Bloomington, IN 47405, USA}
\author{W. G. Newton}
\affiliation{Department of Physics and Astronomy, Texas A\&M University-Commerce, Commerce, TX 75429, USA}
\author{E. O'Connor}
\affiliation{Department of Physics, North Carolina State University, Raleigh, NC 27695, USA; Hubble  Fellow}
\author{L. F. Roberts}
\affiliation{National Superconducting Cyclotron Laboratory and Department of
Physics and Astronomy, Michigan State
University, East Lansing, Michigan 48824, USA}
\date{\today}
\begin{abstract}

Nuclear pasta, with nucleons arranged into tubes, sheets, or other complex shapes, is expected in core collapse supernovae (SNe) at just below nuclear density. We calculate the additional opacity from neutrino-pasta coherent scattering using molecular dynamics simulations. We approximately include this opacity in simulations of SNe. We find that pasta slows neutrino diffusion and greatly increases the neutrino signal at late times of 10 or more seconds after stellar core collapse. This signal, for a galactic SN, should be clearly visible in large detectors such as Super-Kamiokande.
\end{abstract}
\maketitle

About 20 neutrinos were detected from supernova SN\,1987A
\cite{PhysRevD.70.043006} confirming the radiation of $\approx 3\times 10^{53}$
ergs of gravitational binding energy gained during the stellar core collapse.
Several thousands of events are expected from the next galactic core collapse
supernova (SN) \cite{Scholberg2012}.  These neutrinos carry information about
the incompletely known explosion mechanism \cite{Lund:2010, Brandt:2011, Tamborra:2013} and on nucleosynthesis in material
ejected from the newly formed neutron star (or proto-neutron star PNS) by the
intense neutrino flux \cite{Qian:1993, Horowitz:1999wy, Horowitz:2001yv,
Hudepohl:10, Fischer:10, Martinez-Pinedo:2012, Roberts:12b}.
SN  neutrinos may undergo rich, possibly nonlinear, flavor oscillations
\cite{SNoscillations,2015PhRvD..91f5016W}.  Finally the neutrino signal, in
concert with gravitational-wave \cite{PhysRevD.93.042002} and electromagnetic
observations, may provide a historic multi-messenger data set. 

During the SN explosion, the PNS deleptonizes and cools via the emission of neutrinos of all flavors.  The associated neutrino signal at late times may be sensitive to PNS convection \cite{Roberts:12}, neutrino interactions at high densities \cite{Reddy:99,PhysRevC.59.510}, and the possibility of matter falling back onto the PNS.  Finally, the neutrino signal will be dramatically modified if the star collapses to a black hole.   New neutrino detectors \cite{Scholberg2012} such as the Deep Underground Neutrino Experiment (DUNE) \cite{Ankowski:2016lab,Goodman:2015gmv} and Hyper-Kamiokande \cite{HyperK,Migenda:2016xnc} offer exciting capabilities.  DUNE should provide detailed information on electron neutrinos while Hyper-K, with its very large volume, should be particularly good at observing SN neutrinos at late times.

In the PNS interior, densities in excess of nuclear saturation density $\approx 3\times 10^{14}$ g/cm$^3$ are reached.  At just below nuclear density, nuclei are expected to rearrange into nuclear pasta structures with complex tube (spaghetti), sheet (lasagne), or other shapes \cite{PhysRevLett.50.2066}.   These shapes arise because of the competition between short-range nuclear attraction and long-range Coulomb repulsion.  SN neutrinos, with 10s of MeV energies, have wavelengths comparable to the size of these shapes and will scatter coherently from all of the nucleons inside \cite{Horowitz:2004yf,Horowitz:2004pv}.  This scattering may increase the opacity and significantly impact the neutrino signal from core collapse SN.

Recently there have been several calculations of nuclear pasta
structure, see for example \cite{ Schuetrumpf:2016kzq, PhysRevC.93.065806, PhysRevC.94.025806, PhysRevC.93.055801, PhysRevC.92.045806, Furtado:2015vga, PhysRevC.90.055802, Watanabe2012, PhysRevC.83.035803}. There are interesting similarities between spiral shapes that might be present in both nuclear pasta and
biological membranes, despite the two systems differing
in density by 14 orders of magnitude \cite{PhysRevC.94.055801}, see also  \cite{Caplan:2016uvu}.
Observational signatures of pasta in older neutron stars
have recently been discussed: if disordered, pasta could cause magnetic field
decay  \cite{Pons2013} and slow the cooling of neutron star crusts
\cite{PhysRevLett.114.031102, Deibel:2016vbc}. 


In this letter we demonstrate a new observational signature of pasta in PNSs.  We calculate neutrino-pasta scattering cross sections by performing large-scale molecular dynamics simulations of nuclear pasta.  We approximately include these cross sections in astrophysical simulations of core-collapse and PNS cooling.  For the first time, we predict the impact of pasta on neutrino luminosities and mean energies.  Finally, we predict the signal, from a galactic SN in a large water detector such as Super-Kamiokande \cite{Walter:2008ys}.  We find that the signal could be significantly enhanced at late times of 10 or more seconds after stellar core collapse, as nuclear pasta tends to slow neutrino diffusion.

Neutrino-nucleon elastic scattering is modified in the medium by correlations and structure formation.  We write the transport cross section per nucleon in the medium $\sigma$ as $\sigma = \sigma_0 S_{\rm tot}$ with $\sigma_0$ the free transport cross section.  The total response $S_{\rm tot}$ describes the medium modifications.  It depends on the vector response $S_V$, that describes the response of the system to vector currents, and the axial response $S_A$, that describes the response to axial (or spin) currents, see for example Eq. 31 of Ref. \cite{axialresponse},
\begin{equation}
S_{\rm tot}=(1+\xi f_{\rm pasta})= \frac{(1-Y_p)S_V + 5g_a^2 S_A}{ 1-Y_p  + 5g_a^2  }\, .
\label{eq:xi}
\end{equation}
Here $g_a=-1.26$ and the proton fraction is $Y_p$.  For later use, we have also
defined $\xi f_{\rm pasta}$ as the enhancement in $S_{\rm tot}$, from pasta, over the
free response $S_V=S_A=1$. Here $\xi$ is a constant encoding the maximum
enhancement and $0 \leq f_{\rm pasta}(n_b, T, Y_p) \leq 1$ encodes the density $n_b$,
temperature $T$, and proton fraction $Y_p$ region over which pasta is present.

\begin{figure}[ht] 
  \includegraphics[width=1.\columnwidth]{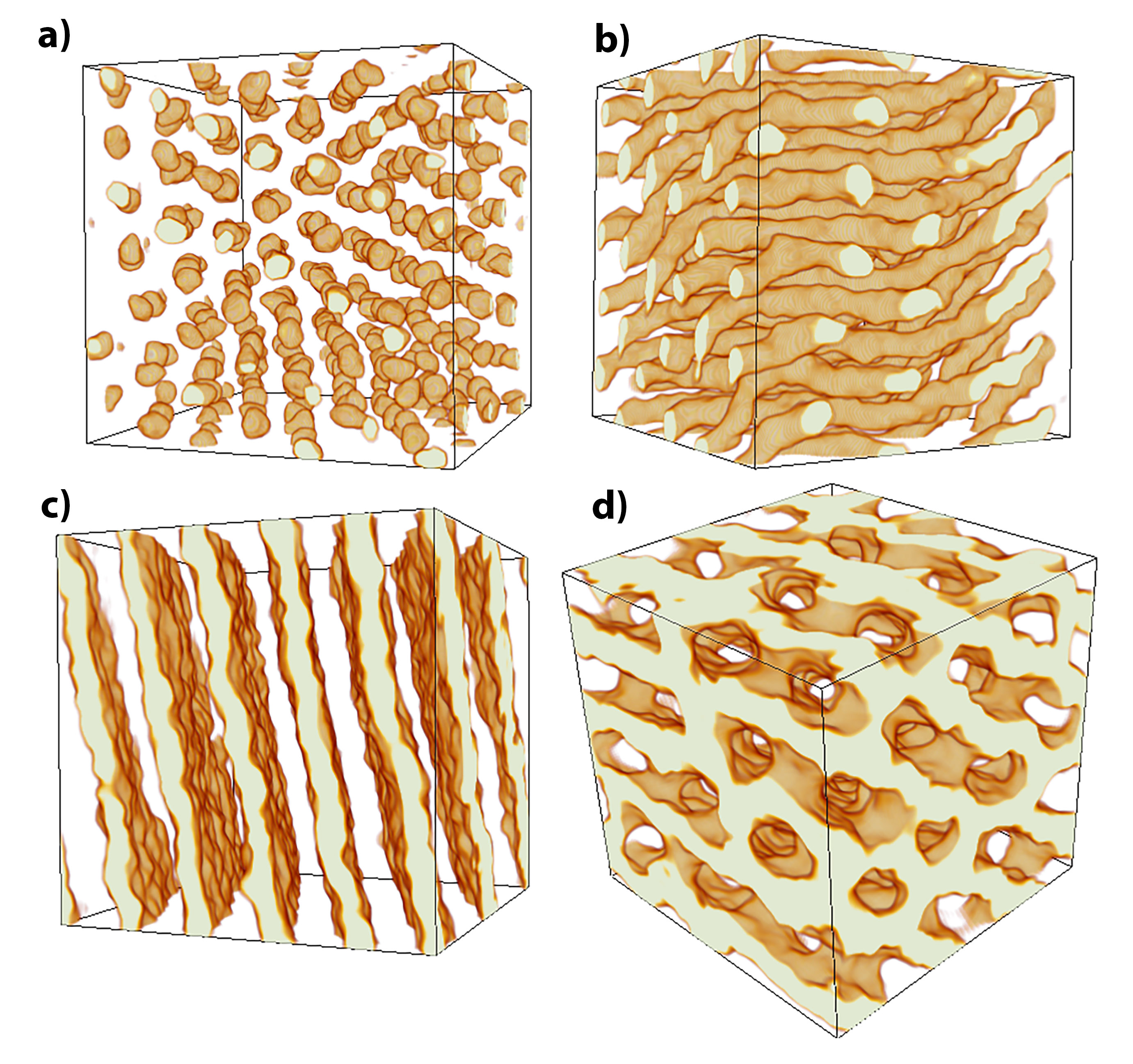}
  \caption{(color on line) Sample configurations from MD simulations with 51,200 nucleons, that show the following shapes, at densities of a) 0.01(isolated nuclei), b) 0.025 (tubes), c) 0.05 (sheets) , and d) 0.075 fm$^{-3}$ (cylindrical holes).}\label{fig:shapes}
\end{figure}

The vector response $S_V$ is an appropriate integral of the neutron static structure factor $S_n(q)$ over momentum transfer $q$ \cite{Horowitz:2004yf},
\begin{equation}
S_V(E_\nu)=\frac{3}{4}\int_{-1}^{1}d\cos\theta\, (1-\cos^2\theta)\, S_n(q)\, .
\label{eq:S_Eave}
\end{equation}
Here $q$ depends on the neutrino energy $E_\nu$ and scattering angle $\theta$, $q^2=2E_\nu^2(1-\cos\theta)$.  
For complex nuclear pasta phases, we calculate $S_n(q)$ using (semi)classical molecular dynamics (MD) simulations,
\begin{equation}
S_n(q) = \langle \rho(q)^*\rho(q) \rangle.
\label{eq.Sn}
\end{equation}
Here the neutron density in momentum space is $\rho(q)=N^{-1/2}\sum_{i=1}^N \exp[i{\bf q}\cdot {\bf r}_i]$ with $r_i(t)$ the location of the ith neutron, $N$ the total number of neutrons in the simulation, and the average in Eq. \ref{eq.Sn} is taken over simulation time $t$.  Sample configurations from our MD simulations are shown in Fig. \ref{fig:shapes}.  We use the model in ref. \cite{Horowitz:2004yf}. Sonoda et al. \cite{PhysRevC.75.042801} have also calculated $S_V(E_\nu)$ in a slightly different QMD model, although for smaller systems with possibly larger finite size effects. They find qualitatively similar results.

\begin{figure}[ht] 
  \includegraphics[width=1\columnwidth]{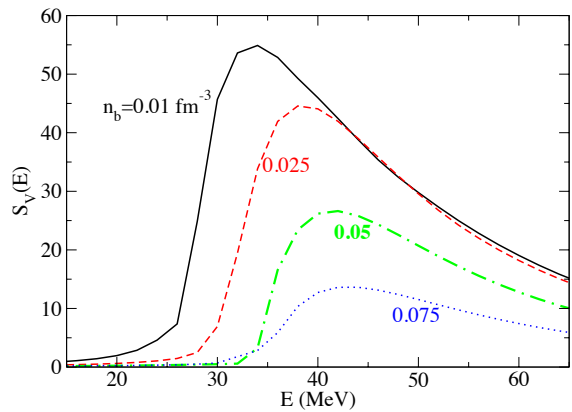}
  \caption{(color on line) Molecular dynamics simulation results for the static structure factor $S_V(E)$ averaged over scattering angles, see Eq. \ref{eq:S_Eave}, vs neutrino energy $E$.  Curves are based on simulations with 51200 nucleons and are at densities of 0.01 (solid), 0.025 (dashed) , 0.05 (dot dashed) and 0.075 fm$^{-3}$ (dotted).}\label{fig:S_Eave}
\end{figure}

Figure \ref{fig:S_Eave} shows $S_V(E_\nu)$, Eq. \ref{eq:S_Eave}, calculated from our MD simulations.  Our MD model has been used to study nuclear pasta in many previous works \cite{Horowitz:2004yf, Horowitz:2004pv, PhysRevC.93.065806, Horowitz:2008vf,  PhysRevC.91.065802,  PhysRevC.90.055805,  Horowitz:2005zb, PhysRevC.94.055801}.  An initial random configuration, at a density $n=0.12$ fm$^{-3}$, is evolved to the desired density by slowly expanding the simulation volume at a rate of $10^{-8}$ c/fm.  Next the system is equilibrated for at least $9\times10^7$ fm/c at constant density, using an MD time step of 2 fm/c.  Then $S_n(q)$ is calculated as a time average of at least 180,000 configurations, each separated by 1,000 fm/c.   This involves an additional time evolution of at least $1.8\times 10^8$ fm/c.  An advantage of this molecular dynamics simulation is that $S_n(q)$ can be directly determined.  Unfortunately this model is likely most accurate at low temperatures and high proton fractions.  Therefore, the above simulations were done at a temperature of $T=1$ MeV and a proton fraction $Y_p=0.4$.  We expect results for other temperatures and proton fractions to be qualitatively similar as long as the basic sizes of the pasta pieces are similar. 

We see that $S_V(E_\nu)$ in Fig. \ref{fig:S_Eave} can be as large as $\approx 50$.  In contrast, the axial response $S_A$ is not expected to be enhanced as the nucleon spins do not add coherently, see for example \cite{axialresponse}.  We are not aware of any explicit calculations of $S_A$ for pasta and use $S_A\approx 1$ as a simple estimate.  This leads to an enhancement for $\xi$, Eq. \ref{eq:xi}, that could be as large as $\xi\approx 5$.  

Alternatively, we consider equations of state that include ``spherical pasta'' phases.  Here, neutrino pasta scattering is modeled as neutrino nucleus elastic scattering from exotic neutron rich heavy nuclei that are assumed nearly spherical in shape but have strong Coulomb correlations between nuclei such that the effective static structure factor is \cite{Horowitz:2004pv},
\begin{equation}
S_n(q) \approx N S_{\rm ion} (q) F(q)^2\, .
\end{equation}
Here $N$ is the neutron number of the heavy nucleus, $F(q)$ is the nuclear elastic form factor and $S_{\rm ion}(q)$ is the static structure factor for charged ions \cite{Sion}.  This describes the correlations between nuclei at long wavelengths. These exotic heavy nuclei are expected to be present for PNS cooling conditions $\approx 3$ or more seconds after stellar core collapse, see below.  For the HS(DD2) equation of state \cite{HSDD2}, and using H. Shen's nuclear composition \cite{1998NuPhA.637..435S}, we find that neutrino scattering from these exotic heavy nuclei, calculated as in ref. \cite{1993ApJ...405..637M}, can increase the neutral current scattering opacity by up to a factor of $\xi\approx 5$ to 10.   This enhancement is consistent with our estimate from MD simulations.

To determine the range of densities $n_b$ and temperatures $T$ where pasta may be present, we have performed fully quantum calculations using the NRAPR Skyrme functional \cite{Steiner2005325}.  These calculations are described in Ref. \cite{PhysRevC.79.055801, PhysRevC.90.065802}.  At $Y_p=0.3$, we find pasta phases for $n_b=0.03$ to 0.11 fm$^{-3}$ and up to a maximum $T=11$ MeV.  This region slightly decreases with decreasing $Y_p$.  For $Y_p=0.1$, pasta is present for $n_b=0.04$ to 0.09 fm$^{-3}$ and up to $T=7$ MeV.  At lower densities, large spherical nuclei will likely be present that will also increase the scattering.  Therefore, the region with an enhanced neutrino opacity (compared to free nucleons) probably extends to somewhat lower densities.   This is shown in Fig. \ref{fig:shapes} a) and Fig. \ref{fig:S_Eave} where the $n_b=0.01$ fm$^{-3}$ MD simulation actually involves nucleons clustered into isolated nuclei.

\begin{figure}[ht] \centering
 \includegraphics[width=0.9\columnwidth]{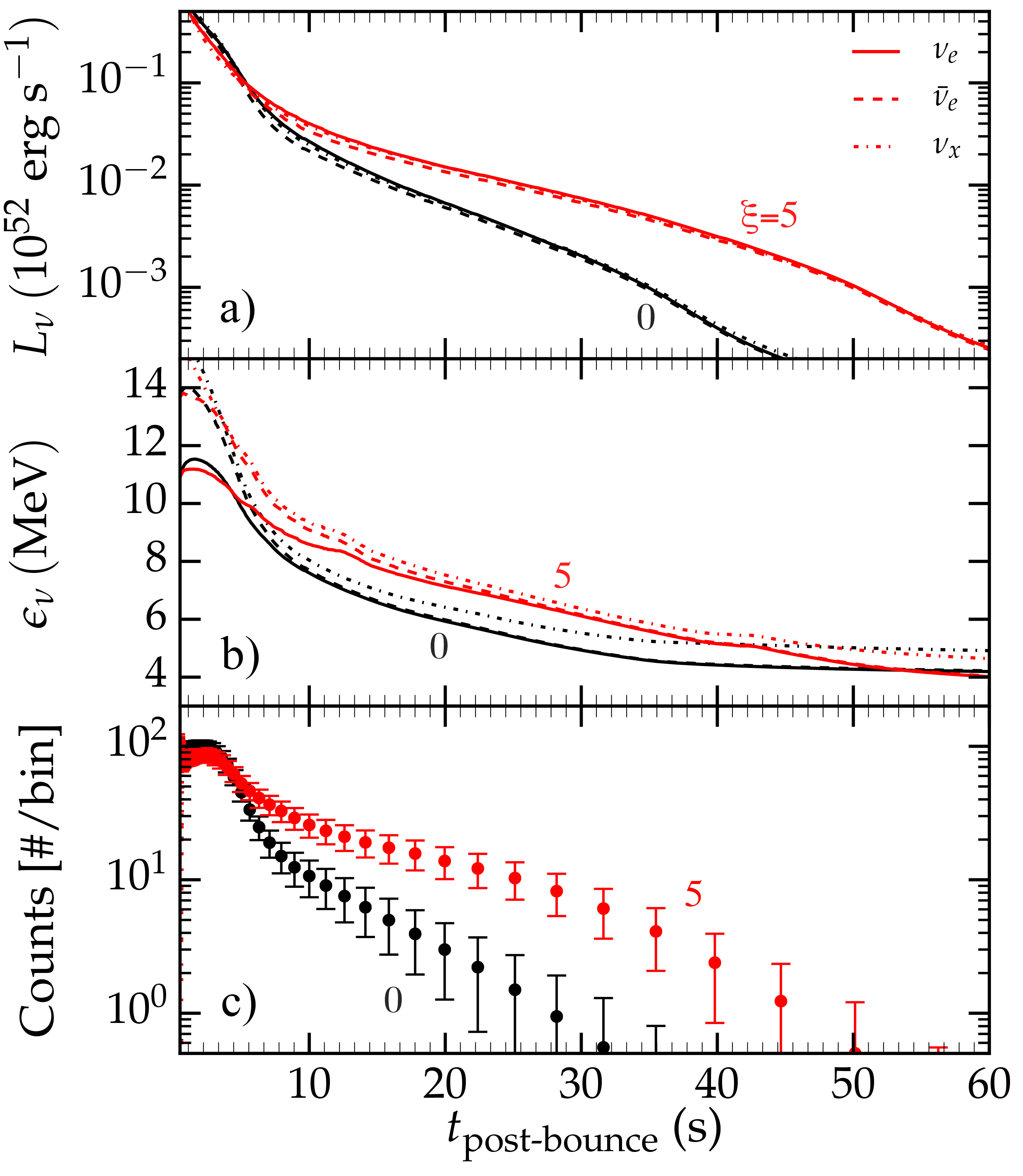}
  \caption{(Color on line) Neutrino luminosity a) (top), mean neutrino energy b) (middle), and
  approximate number of Super-Kamiokande counts at 10 kpc c) (bottom) versus time since core
  bounce. Shown is the  baseline simulation with no opacity enhancement
  ($\xi=0$, black), and one simulation using
  $\xi=5$ (red). The counts are shown for logarithmically spaced
  time bins and the error bars are Poissonian. The presence of pasta enhances
  the neutrino luminosity in all flavors at late times and also increases the
  average energies of all flavors of neutrinos. These two effects combine to
  significantly increase the count rate at late times for models including pasta.}
  \label{fig:nulum} 
\end{figure}

\begin{figure}[ht] \centering
 \includegraphics[width=0.9\columnwidth]{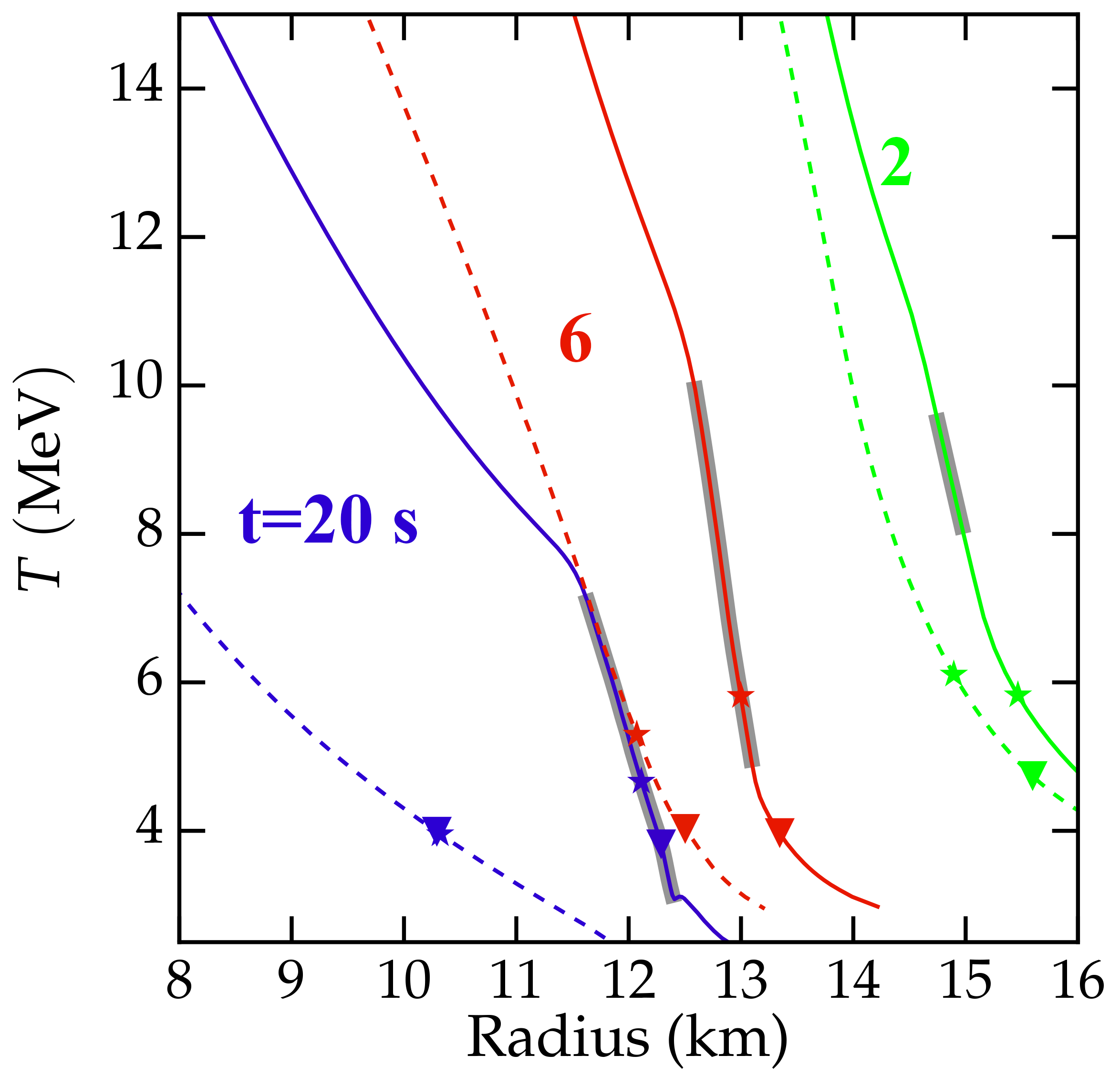}
  \caption{(Color on line) Temperature structure of the PNS atmosphere versus radius with (solid lines, $\xi=5$) and
  without (dashed lines, $\xi=0$) parameterized pasta. The pasta is present in the gray
  regions. The green, red, and blue lines are for times $t=2$, 6, and 20 s after
 core collapse, respectively. The triangles and stars correspond to the electron
  neutrino and anti-neutrino spheres, respectively. The critical temperature of
  the pasta prevents the inner layers of the atmosphere from reducing their
  temperature, resulting in a hotter, extended atmosphere. This results in the
  larger luminosities and neutrino average energies in models including pasta. }
  \label{fig:surface} 
\end{figure}

We explore two phases of core collapse SN where pasta may
impact the thermodynamics and neutrino emission. 
Pasta is included parametrically by enhancing the neutrino scattering opacity in
regions where we expect the pasta to be present. This region is defined by
$f_{\textrm{pasta}} = h(n_b/n_{\textrm{min}}-1, 0.3)h(1 - n_b/n_{\textrm{max}},
0.1)h(1-T/T_{\textrm{crit}}, 0.1),$ where $h(x,y) = 1/2 + 1/2 \tanh(x/y)$. The
scattering opacity is then corrected to be $\kappa_s = \kappa_{s,0}(1 + \xi 
f_{\textrm{pasta}})$, where $\kappa_{s,0}$ is the scattering opacity in
the absence of pasta. Based on the results of our pasta calculations, we choose
$n_{\textrm{min}} = 0.01 \, \textrm{fm}^{-3}$, $n_{\textrm{max}} = 0.1 \,
\textrm{fm}^{-3}$, and $T_{\rm crit} = 10 \, \textrm{MeV}$. We expect that
changing these parameters will impact the details of how pasta impacts
our simulations, but these suffice for exploratory calculations.

The first presence of pasta occurs late in the collapse phase, $\sim
1\,$ms prior to core bounce (when nuclear density is reached). We explore this phase with one-dimensional collapse
simulations using {\tt{GR1D}} \cite{GR1D}. The pasta is located well
inside the electron neutrino decoupling radius and since these
neutrinos make up the bulk of the neutrino energy density at this
time, we see no dynamic or thermodynamic effect of the presence of
pasta.  However, there is a perceptible, yet very small impact on the
$\nu_x$ signal at this time.  The decoupling density for $\nu_x$, for
a very brief time, reaches upwards of $\sim 8 \times
10^{12}$\,g\,cm$^{-3}$. With the presence of pasta, we see a
suppression in the $\nu_x$ luminosities of $\sim 3-4$ times, in the
millisecond preceding bounce. Since the $\nu_x$ luminosity at this
point is only $\sim 10^{49}$\,erg\,s$^{-1}$ and the average neutrino
energy is low, we do not expect this effect to be observable with
current or even next generation detectors.

Although the impact of pasta on the infall phase neutrino emission is very small, it may
significantly alter the late time neutrino cooling signal when pasta forms in
the atmosphere of the PNS. Using a variant of the spherically symmetric
radiation hydrodynamics code described in \cite{Roberts:12a}, we evolve a $15 \,
M_\odot$ progenitor \cite{Woosley:02} from core collapse through the accretion
phase. Once the shock has passed a baryonic mass coordinate of $1.5 \, M_\odot$,
we excise the outer layers of the star and evolve only the PNS over 100 s.
Convection is included as in \citep{Roberts:12} through a mixing length theory
prescription. We employ a new implementation of the LS220 EoS \cite{Lattimer:91, Schneider:16} and we use the simple
neutrino opacities of \cite{Bruenn:85}.  We expect more realistic opacities for
homogeneous material to change the cooling timescale, but not qualitatively
impact the difference between models with and without pasta \citep{Reddy:99, 
Hudepohl:10, Roberts:12}. 

 
Here we present two models, one without parameterized pasta opacities $\xi=0$ and one
with parameterized pasta opacities $\xi=5$. In Fig. \ref{fig:nulum}, the neutrino
emission properties of these two models are shown. Pasta enhances the late time
neutrino luminosity and the average energies of all neutrino species are
increased. This is due to the impact of pasta on the PNS atmosphere, where pasta
forms at late times. The temperature structure of the atmosphere at select times
is shown in Fig. \ref{fig:surface}. The pasta is present at one second after
bounce near the surface of the PNS and persists until late times. Rather than
cooling to a low temperature, the interior of the PNS tries to stay above the
pasta critical temperature to maintain a low opacity and increase the rate of
energy transport. This results in a stronger temperature gradient near the PNS
surface, a higher temperature of neutrino decoupling, and a more radially
extended atmosphere. The higher decoupling temperature increases the neutrino
average energies, while the larger radius and increased decoupling
temperature both serve to increase the neutrino luminosities. We expect the results
to be sensitive to the critical temperature. If $T_\textrm{crit}$ is smaller
than we have assumed, we expect a smaller impact on the neutrino emission. 

In the bottom panel of Fig. \ref{fig:nulum} we show the expected count rates for
these models in Super-Kamiokande calculated using the {\tt{SNOwGLObES}} package
\cite{Scholberg2012}. We assume a distance of 10 kpc, a detector mass of 32 kT, and for simplicity only include inverse beta decay events from $\bar\nu_e$.  The total number of events is about 3400.  Clearly, the pasta has a strong impact on the late time detection
prospects and can enhance the count rate by almost an order of magnitude. For
the two particular models considered here, only $51 \pm 7$ counts are expected
after 10 seconds for the model without pasta while $181 \pm 13$ counts are
expected for the model including pasta.  Both the enhanced luminosity and the
larger average energies increase detection efficiency. Therefore, a signature of
the nuclear pasta should be visible in the late-time neutrino signal of a
galactic CCSN. 

Future work should improve the calculation of neutrino-pasta scattering and how
this, and its energy dependence, is incorporated into proto-neutron star cooling simulations.  In addition,
the late time neutrino signal should be studied in more detail for large
detectors such as Super-K, Hyper-K, or possibly Ice Cube \cite{Icecube}.

In conclusion, we have used molecular dynamics simulations to calculate neutrino-pasta scattering and approximately included this in simulations of proto-neutron star cooling.  We find that nuclear pasta can slow the diffusion of neutrinos and significantly enhance the late time neutrino signal at times of 10 to 100 seconds after stellar core collapse.  This signal, for a galactic SN, should be clearly visible in large neutrino detectors such as Super-Kamiokande, and its observation could provide information on neutrinos scattering from nuclear pasta shapes. 

We thank Shirley Li and John Beacom for helpful discussions and the Institute for Nuclear Theory at the University of Washington where some of this work was done.  This work was supported in part by DOE Grants DE-FG02-87ER40365 (Indiana
University) and DE-SC0008808 (NUCLEI SciDAC Collaboration).  Support for
this work was provided also by NASA through Hubble Fellowship Grant
\#51344.001-A awarded by the Space Telescope Science Institute, which
is operated by the Association of Universities for Research in
Astronomy, Inc., for NASA, under contract NAS 5-26555.  TF acknowledges support from the Polish National Science Center (NCN) under grant number UMO-2011/02/A/ST2/00306.  WN was supported by the Research Corporation for Science Advancement through the Cottrell College Science Award \#22741. This work was enabled in part by the NSF under Grant No. PHY-1430152 (JINA Center for the Evolution of the Elements). Computer time was provided by the INCITE program. This research used resources of the Oak Ridge Leadership Computing Facility located at Oak Ridge National Laboratory, which is supported by the Office of Science of the Department of Energy under Contract No. DEAC05-00OR22725. 


\end{document}